\documentclass[prd,a4paper,twocolumn, superscriptaddress,floatfix]{revtex4}
\usepackage{graphicx}
\usepackage{bbm}
\usepackage{amssymb}
\usepackage{hyperref}
\usepackage{multirow}
\begin{document}
\newcommand{\be}{\begin{equation}}
\newcommand{\ee}{\end{equation}}
\newcommand{\bq}{\begin{eqnarray}}
\newcommand{\eq}{\end{eqnarray}}
\newcommand{\bsq}{\begin{subequations}}
\newcommand{\esq}{\end{subequations}}
\newcommand{\bc}{\begin{center}}
\newcommand{\ec}{\end{center}}
\newcommand\lapp{\mathrel{\rlap{\lower4pt\hbox{\hskip1pt$\sim$}} \raise1pt\hbox{$<$}}}
\newcommand\gapp{\mathrel{\rlap{\lower4pt\hbox{\hskip1pt$\sim$}} \raise1pt\hbox{$>$}}}
\newcommand{\dpar}[2]{\frac{\partial #1}{\partial #2}}
\newcommand{\sdp}[2]{\frac{\partial ^2 #1}{\partial #2 ^2}}
\newcommand{\dtot}[2]{\frac{d #1}{d #2}}
\newcommand{\sdt}[2]{\frac{d ^2 #1}{d #2 ^2}}
\newcommand{\vv}{\bar{v}}
\newcommand{\cc}{\tilde c}
\newcommand{\dd}{\tilde d}
\newcommand{\pr}{\mathcal{P}}
\newcommand{\sgwb}{\Omega_{\rm GW}}
\newcommand{\hgw}{h_{\rm GW}}

\title{Probing Cosmic Superstrings with Gravitational Waves}

\author{L. Sousa}
\email[Electronic address: ]{Lara.Sousa@astro.up.pt}
\affiliation{Instituto de Astrof\'{\i}sica e Ci\^encias do Espa{\c c}o, Universidade do Porto, CAUP, Rua das Estrelas, PT4150-762 Porto, Portugal}
\affiliation{Centro de Astrof\'{\i}sica da Universidade do Porto, Rua das Estrelas, PT4150-762 Porto, Portugal}

\author{P.P. Avelino}
\email[Electronic address: ]{pedro.avelino@astro.up.pt}
\affiliation{Instituto de Astrof\'{\i}sica e Ci\^encias do Espa{\c c}o, Universidade do Porto, CAUP, Rua das Estrelas, PT4150-762 Porto, Portugal}
\affiliation{Centro de Astrof\'{\i}sica da Universidade do Porto, Rua das Estrelas, PT4150-762 Porto, Portugal}
\affiliation{Departamento de F\'{\i}sica e Astronomia, Faculdade de Ci\^encias, Universidade do Porto, Rua do Campo Alegre 687, PT4169-007 Porto, Portugal}

\begin{abstract}
We compute the stochastic gravitational wave background generated by cosmic superstrings using a semi-analytical velocity-dependent model to describe their dynamics. We show that heavier string types may leave distinctive signatures on the stochastic gravitational wave background spectrum  within the reach of present and upcoming gravitational wave detectors. We examine the physically motivated scenario in which the physical size of loops is determined by the gravitational backreaction scale and use NANOGRAV data to derive a conservative constraint of $G\mu_F<3.2 \times 10^{-9}$ on the tension of fundamental strings. We demonstrate that approximating the gravitational wave spectrum generated by cosmic superstring networks using the spectrum generated by ordinary cosmic strings with reduced intercommuting probability (which is often done in the literature) leads, in general, to weaker observational constraints on $G\mu_F$. We show that the inclusion of heavier string types is required for a more accurate characterization of the region of the $(g_s,G\mu_F)$ parameter space that may be probed using direct gravitational wave detectors. In particular, we consider the observational constraints that result from NANOGRAV data and show that heavier strings generate a secondary exclusion region of parameter space.
\end{abstract}

\maketitle
\flushbottom

\section{Introduction}

The recent detection of gravitational waves by the Laser Interferometer Gravitational-wave Observatory (LIGO) \cite{Abbott:2016blz} ushered a new era of astronomy and fundamental physics by opening the possibility of observing previously undetected or poorly known sources (see \cite{Sathyaprakash:2009xs,Marx:2011hf} for recent reviews). One such source are $1+1$-dimensional topological defects known as cosmic strings. The production of cosmic string networks as remnants of symmetry breaking phase transitions is expected in a large variety of grand unified scenarios \cite{vilenkin2000cosmic}.  Moreover, recent developments in string theory raised the possibility that fundamental strings (F-strings) and $1$-dimensional Dirichlet branes (D-strings) may grow to macroscopic sizes and play the role of cosmic strings. These F- and D-strings --- generally referred to as cosmic superstrings --- are expected to have reduced intercommuting probabilities and to form highly entangled networks with junctions that have an hierarchy of tensions \cite{Copeland:2003bj,Jackson:2004zg}. Cosmic superstrings are expected to be copiously created at the end of several brane inflationary scenarios (which often end with a symmetry breaking phase transition) \cite{Sarangi:2002yt,Jones:2002cv,Majumdar:2002hy}. Although cosmic strings and  cosmic superstrings are created in the early universe, they are expected to persist and survive throughout cosmic history. However --- in spite of their characteristic imprints on the cosmic microwave background (CMB) \cite{Pogosian:1999np,Ade:2013xla,Lizarraga:2014xza,Lazanu:2014xxa}, small structure formation \cite{Wu:1998mr,Pogosian:2008am,Lin:2015asa}, reionization history \cite{Avelino:2003nn,Avelino:2004wm,Olum:2006at} and gravitational lensing observations \cite{deLaix:1997jt,Mack:2007ae,Thomas:2009bm,Yamauchi:2011cu} --- both cosmic strings and cosmic superstrings have as of yet evaded detection.

Cosmic string interactions play a key role in the evolution of cosmic string networks. They often result in the formation of cosmic string loops that detach from the network and decay by emitting gravitational waves. The superposition of the transient gravitational wave bursts emitted by cosmic string loops --- which exist in large numbers at any given time in cosmic history --- is expected to generate a stochastic gravitational wave background \cite{Vilenkin:1981bx,Hogan:1984is,Brandenberger:1986xn,Accetta:1988bg}. Different regions of the gravitational wave spectrum may be probed using current and upcoming astrophysical experiments: direct gravitational wave detectors --- either ground-based (Advanced LIGO \cite{Sigg:2008zz}, Advanced Virgo \cite{Accadia:2011zzc} and KAGRA \cite{Kuroda:2010zzb}) or space-borne (evolved LISA\cite{AmaroSeoane:2012km} and DECIGO \cite{Kawamura:2011zz}); pulsar timing experiments (Parkes \cite{Manchester:2007mx}, European \cite{Ferdman:2010xq} pulsar timing arrays, NANOGRAV \cite{Demorest:2012bv} and the Square Kilometer Array \cite{Janssen:2014dka}); small-scale fluctuations and B-mode polarization of the CMB \cite{Smith:2006nka}; and big bang nucleosynthesis \cite{Cyburt:2004yc}. There is thus the prospect either for the detection of the stochastic gravitational wave background generated by cosmic string and superstring networks or for the tightening of current constraints on their tension.

The stochastic gravitational wave background generated by ordinary cosmic string networks has been extensively studied in the literature \cite{Caldwell:1991jj,Caldwell:1996en,Hogan:2006we,Siemens:2006yp,DePies:2007bm,Olmez:2010bi,Binetruy:2012ze,Sanidas:2012ee,Kuroyanagi:2012wm,Kuroyanagi:2012jf,Sanidas:2012tf,Sousa:2013aaa,Sousa:2014gka,Kuroyanagi:2016ugi}. This is, however, not the case for cosmic superstrings. Some attempts at modeling the spectrum generated by these strings have been made (see e.g. \cite{Siemens:2006yp,Binetruy:2010cc,Sanidas:2012tf}). These often treat cosmic superstrings as ordinary cosmic strings with reduced intercommuting probability, considering only the lightest strings (F-strings) and ignoring the effect of junctions on the network's dynamics. As demonstrated in \cite{Sousa:2013aaa}, an accurate modeling of the large scale dynamics of the networks is essential for an accurate characterization of the shape and amplitude of the gravitational wave spectrum generated by them. Here, we compute the stochastic gravitational wave background generated by cosmic superstring by resorting to the semi-analytical velocity-dependent model for cosmic superstrings described in \cite{Avgoustidis:2007aa}. This model allows us, not only to incorporate the effect of the junctions, but also to compute the contribution of heavier string types to the spectrum. We find that, although in some instances the lightest superstring type may be well described using ordinary cosmic strings with reduced intercommuting probability to model their dynamics, heavier modes may leave observable signatures in the stochastic gravitational wave spectrum. We thus show that including these heavier strings is essential to accurately characterize the shape and amplitude of the spectrum.

This paper is organized as follows. In Sec. \ref{stringevo}, we briefly review the velocity-dependent one-scale model for cosmic string network dynamics. In Sec. \ref{sgwb}, we describe the emission of gravitational waves by cosmic string loops, and the method for computing the stochastic gravitational wave background they generate. In Sec. \ref{superstrings}, we review the velocity-dependent one-scale model for cosmic superstrings. In Sec. \ref{sgwbsuper}, we start by investigating the impact of the junctions on the dynamics of F-strings (Sec. \ref{voscomp}). We, then, study the effect of including heavier string types on the shape of the stochastic gravitational wave background spectrum, in Sec \ref{heavier}, for different loop sizes. In Sec. \ref{realistic}, we discuss the spectrum of gravitational waves generated by realistic cosmic superstring networks. In Sec. \ref{obs}, we derive the observational constraints on the tension of fundamental strings, $G\mu_F$, for which loop size is determined by the gravitational backreaction scale, set by (indirect) CMB data (Sec. \ref{cmbsec}) and NANOGRAV (Sec. \ref{pulsarsec}). We the conclude in Sec. \ref{conc}.

\section{Cosmic String Evolution\label{stringevo}}

The Velocity-dependent One-Scale (VOS) model \cite{Martins:1996jp,Martins:2000cs} describes the time evolution of the characteristic lengthscale of the network, $L$, and of its root-mean-square (RMS) velocity, $\vv$, thus allowing for a quantitative characterization of string network dynamics. If one assumes that the cosmic string network is roughly homogeneous on sufficiently large scales, one may define its characteristic lengthscale as

\be
\rho=\frac{\mu}{L^2}\,,
\ee
where $\mu$ is the cosmic string tension, and $\rho$ is the average energy density of long strings. For infinitely thin cosmic strings --- whose thickness is much smaller than their curvature radius --- the following evolution equations for $L$ and $\vv$ can be obtained by averaging the microscopic Nambu-Goto equations of motion \cite{Martins:1996jp,Martins:2000cs} (see also \cite{Avelino:2011ev,Sousa:2011ew,Sousa:2011iu} for a more general derivation of the VOS equations):

\bq
2\frac{dL}{dt} & = & \left(2H+\frac{\vv^2}{\ell_d}\right)L\,,\label{vosL}\\
\frac{d\vv}{dt} & = & (1-\vv^2)\left[\frac{k(\vv)}{L}-\frac{\vv}{\ell_d}\right]\,,\label{vosv}
\eq
where $H=(da/dt)/a$ is the Hubble parameter, and $a$ is the cosmological scale factor. We have also introduced the damping lengthscale, $\ell_d^{-1}=2H+\ell_f^{-1}$, that accounts not only for the deceleration caused by the Hubble expansion but also for the effect of frictional forces caused by interactions with other fields (encoded in the frictional lengthscale, $\ell_f$). We shall assume for the remainder of this paper that $\ell_f=\infty$. Moreover, $k(\vv)$ is an adimensional curvature parameter that encodes the effects caused by the existence of small-scale structure on long strings. In \cite{Martins:2000cs}, the following ansatz was suggested
\be
k(\vv)=\frac{2\sqrt{2}}{\pi}\left(1-\vv^2\right)\left(1+2\sqrt{2}\vv^3\right)\frac{1-8\vv^6}{1+8\vv^6}\,.
\label{kpar}
\ee

String interactions are a key ingredient in the evolution of cosmic string networks. When two cosmic strings collide, they may exchange partners and intercommute. This process has important consequences for the dynamics of a network, since it leads to the production of cosmic string loops that detach from the long-string network. The energy lost into loops by the long-string network may be written as \cite{Kibble:1984hp}

\be
\left.\frac{d\rho}{dt}\right|_{\rm loops}={\tilde c}{\bar v}\frac{\rho}{L}\,,
\label{loss}
\ee
where ${\tilde c}$ is a phenomenological parameter that characterizes the efficiency of the loop-chopping mechanism. Numerical simulations indicate that ${\tilde c}=0.23\pm 0.04$ is a good fit both in the matter and radiation eras, for standard cosmic string networks \cite{Martins:2000cs}.

These loops start decaying radiatively once they detach from the cosmic string network and, thus, they have a finite lifespan. Consequently, the network loses energy at the rate given by Eq. (\ref{loss}). This effect is included in the VOS equations by adding the following term to the right-hand side of Eq. (\ref{vosL}):

\be
\left.\frac{dL}{dt}\right|_{loops}=\frac{1}{2}\tilde{c}\vv\,.
\label{lossL}
\ee

Eqs. (\ref{vosL}), (\ref{vosv}) and (\ref{lossL}) are the basis of the VOS model and they describe the large-scale evolution of cosmic string networks. Interestingly, the linear scaling regime \cite{Bennett:1987vf,Albrecht:1989mk,Allen:1990tv,Copeland:1991kz,Vincent:1996rb} arises naturally in this model. Indeed, a regime of the form 
\be
\frac{L}{t}=\xi= \sqrt{\frac{k(k+{\tilde c})}{4\beta(1-\beta)}}\,\qquad\mbox{and}\qquad\vv=\sqrt{\frac{k}{k+{\tilde c}}\frac{1-\beta}{\beta}}\,,
\label{sca-def}
\ee
is an attractor solution of the VOS equations, in the case of a decelerating power-law expansion of the universe --- with $a\propto t^\beta$ and $0<\beta<1$. (For a detailed discussion of this and other scaling solutions of cosmic string and other p-brane networks, see \cite{Sousa:2011ew,Sousa:2011iu,Avelino:2012qy}.) Note however that, in a realistic cosmic background, such scaling solutions are only attainable deep into the radiation and matter epochs (for $\beta=1/2$ and $\beta=2/3$, respectively). During the radiation-matter transition, the network enters a long-lasting transitional period during which it is not in a linear scaling regime \cite{Avelino:2012qy,Sousa:2013aaa}. This `delay' in scaling is what one should  realistically  expect: cosmic string networks have to slowly adapt, after the onset of the radiation-matter transition, to the changes in the underlying cosmological background. Note also that the matter era might not be long enough for the network to reestablish scale-invariant evolution before the onset of dark energy domination. In the latter phase, the expansion of the universe is accelerated and the network becomes conformally stretched \cite{Sousa:2011ew} with $L\propto a$ and $\vv\to0$. Cosmic string networks are, then, diluted away rapidly by the accelerated expansion once the universe becomes dark-energy-dominated.

Finally, it should be remarked that the VOS model has enough plasticity to allow for the description of non-standard cosmic string networks, through a recalibration of its free parameters and/or through the inclusion of additional terms (as we shall see in Sec. \ref{superstrings} for the case of cosmic superstrings).

\section{The Stochastic Gravitational Wave Background generated by Cosmic Strings\label{sgwb}}

The creation of cosmic string loops is expected to occur copiously throughout the evolution of cosmic string networks. Once a loop detaches from the long-string network, it oscillates relativistically and decays through the emission of gravitational radiation. There are, at any given time in cosmic history, several cosmic string loops emitting gravitational waves (GWs) in different directions. The superposition of these emissions gives rise to a Stochastic Gravitational Wave Background (SGWB) with a characteristic shape, spanning a wide range of frequencies \cite{Vilenkin:1981bx,Hogan:1984is,Brandenberger:1986xn,Accetta:1988bg}.

\subsection{Cosmic String Loop Emission}

In this context, it is often assumed that cosmic string loops are created with a size that is a fixed fraction of the characteristic length of the network at the time of birth ($t_b$)

\be
l_b=\alpha L(t_b)\,,
\ee
where $\alpha$ is a constant parameter. Realistically, one does not expect all the loops produced at a given time to have precisely the same length. Instead, the distribution of the sizes of the loops formed at the time $t_b$ is expected to have a peak around $l_b$. If the width of the distribution of loop sizes is not very large, assuming that all the loops have the same size at the moment of formation should be a good approximation (the effect of relaxing this assumption was studied in \cite{Sanidas:2012ee}).

Loops emit gravitational waves in a discrete set of frequencies

\be
f_j=\frac{2j}{l}\,,
\label{frdef}
\ee
where $l$ is the length of the loop, $j$ is the harmonic mode number and $f_j$ is the corresponding frequency. Gravitational back-reaction damps higher frequency modes more efficiently than it does low-frequency ones \cite{Battye:1994qa,Battye:1997ji}. The power emitted in each mode is

\be
\frac{dE_j}{dt}=G\mu^2 \frac{\Gamma}{\mathcal{E}}j^{-q}\,,
\ee
where $\mathcal{E}=\sum_m^{n_s}m^{-q}$ and $q$ is a parameter that depends on the shape of the loops. It has been shown that $q \approx 2$ for loops with kinks and $q \approx 4/3$ for cuspy loops \cite{vilenkin2000cosmic}. Here, we have also introduced a cut-off, $n_s$, to the summation in $\mathcal{E}$. Previous work \cite{Sanidas:2012ee,Sousa:2013aaa} has shown that it is sufficient to consider modes up to $n_s=10^3(10^5)$ for loops with kinks (cusps): the spectrum remains unchanged by the inclusion of higher order modes. Moreover, $\Gamma \sim 65$ \cite{Vilenkin:1981bx,Quashnock:1990wv} (see also \cite{Allen:1994bs,Allen:1994ev,Casper:1995ub}) is a parameter characterizing the efficiency of GW emission, and $G$ is the gravitational constant.

Cosmic string loops thus lose energy roughly at a constant rate, $dE/dt=\Gamma G\mu^2\,$, and their length decreases as GWs are emitted:

\be
l(t)=\alpha L(t_b)-\Gamma G\mu (t-t_b)\,,
\label{loopsize}
\ee
for $t_b<t<t_d=\left((\alpha L(t_b))/(\Gamma G\mu)+1\right)^{-1}t_b$, where $t_d$ is the time of the loop disappearance.

There is presently no consensus in the literature regarding the most appropriate choice of $\alpha$. Several early studies \cite{Albrecht:1984xv,Bennett:1989yp,Allen:1990tv,Vincent:1996rb} suggested that $\alpha$ should be smaller than (or close to) the gravitational back-reaction scale $\alpha \lesssim \Gamma G\mu$, while more recent studies \cite{Vanchurin:2005yb,Vanchurin:2005pa,Ringeval:2005kr,Martins:2005es,Olum:2006ix} suggest scales much closer --- albeit 1 to 3 orders of magnitude smaller ---  to the characteristic lengthscale of the network. There are also recent studies which indicate that string loops might be formed with lengths similar to the string thickness \cite{Vincent:1996qr,Vincent:1997cx,Bevis:2007gh}. Recently, however, there is mounting evidence \cite{Vanchurin:2007ee,Lorenz:2010sm,BlancoPillado:2011dq} (coming from Nambu-Goto simulations) that loop formation may happen at two fundamentally different scales, so that two different populations of loops are produced. As a matter of fact, recent simulations \cite{Blanco-Pillado:2013qja} seem to indicate that $90\%$ of the energy lost due to loop production is in the form of small loops with $\alpha \sim \Gamma G\mu$, while the rest is mostly in the form of large loops with $\alpha\sim 1/20$.

\subsection{Spectral density of gravitational waves\label{specden}}

The amplitude of the stochastic gravitational wave background is often measured by the energy density in gravitational waves, $\rho_{\rm GW}$, per logarithmic frequency interval in units of critical density ($\rho_{\rm crit}$),

\be
\Omega_{\rm GW}(f)=\frac{1}{\rho_{{\rm crit}}}\frac{d\rho_{\rm GW}}{d \log f}\,,
\ee
where $\rho_{\rm crit}=3H_0^2/(8\pi G)$ and the subscript `0' denotes the present epoch. This spectral density may be written as \cite{Sousa:2014gka}:

\be
\Omega_{\rm GW}(f)=\sum_j^{n_s}\frac{j^{-q}}{\mathcal{E}}\Omega_{\rm GW}^j(f)\,,
\ee
with

\be
\sgwb^j(f)=\frac{16\pi}{3}\left(\frac{G\mu}{H_0}\right)^2\frac{\Gamma}{f\,a_0^5}\int_{t_i}^{t_0}j n(l_j(t'),t')a^5(t')dt'\,,
\ee
where $t_i$ is the time instant in which the chopping of loops from the long-string network begins, and $n(l,t')dl$ is the number density of cosmic string loops with lengths between $l$ and $l+dl$ at time $t$. Here we have also defined

\be
l_j(t')\equiv\frac{2j}{f}\frac{a(t')}{a_0}\,
\label{loopsize}
\ee
as the physical length that the cosmic string loops should have at each instant $t'$ in order to emit, in the harmonic mode $j$, gravitational waves that have a frequency $f$ at the present time.

It is straightforward to show that

\be
\sgwb^j(jf)=\sgwb^1(f)\,,
\ee
since $l_j(t')$ and $l_1(t')$ are identical if the frequencies under consideration satisfy $f_j=j f_1$. Therefore, once the spectral density of GWs emitted in the fundamental mode ($n=1$), $\sgwb^1(f)$, is computed, one may easily construct $\sgwb^j(f)$ for any arbitrary emission mode $j$ (thus significantly reducing computation time).

\subsection{Number Density of Loops}

In order to compute $\Omega_{\rm GW}(f)$, it is crucial to have a good estimate of the loop distribution function $n(l_j(t'),t')$. Let $n_c(t)$ be the total number density of loops that have been formed as the result of intercommutation between the time of formation of the string network and a time $t$. The VOS model for cosmic strings implies that the rate of loop production per unit volume is

\be
\frac{dn_c}{dt}=\frac{\tilde c}{\alpha}\frac{\vv}{L^4}\,.
\label{nc}
\ee
This expression may simply be obtained by dividing the total energy density that is lost by the string network due to loop formation (Eq. (\ref{loss})) by the energy of each loop at the moment of creation.

After formation, loop size shrinks as a consequence of gravitational radiation emission. Therefore, $n(l_j(t'),t')$ has contributions from all preexisting loops that have physical lengths $l_j(t')$ at time $t'$. Determining the times of creation ($t_b^i$) of the loops  that contribute to a given frequency at any given time $t'$ is essential to computing $n(l_j(t'),t')$. Given these instants, the number density of loops is given \cite{Sousa:2013aaa}

\be
n\left(l_j(t'),t'\right)  =  \sum_i \left\{ \frac{1}{\alpha \left. \frac{dL}{dt}\right|_{t=t_b^i}+\Gamma G\mu} \frac{\tilde c}{\alpha} \frac{\vv(t_b^i)}{L^4(t_b^i)}\left(\frac{a(t_b^i)}{a(t')}\right)^3\right\}\,.
\label{loopdist}
\ee
Note that, given the strong dependency of Eqs. (\ref{nc}) and (\ref{loopdist}) on $L$ and $\vv$, an accurate characterization of the large scale cosmic string dynamics is necessarily for a precise estimation of the number density of loops. So, in performing this calculation --- contrary to what is generally done in the literature (see e. g. \cite{DePies:2007bm,Binetruy:2012ze,Sanidas:2012ee,Kuroyanagi:2012wm}) --- we did not assume the network to be in a linear scaling regime. It was demonstrated in \cite{Sousa:2013aaa} that this assumption leads to an underestimation of the number density of loops produced in the matter era, and thus to a significant underestimation of the amplitude of the SGWB generated by cosmic strings during this epoch.

\subsection{The Small-loop Regime}

In Ref. \cite{Sousa:2014gka}, an alternative method for the computation of the SGWB generated by small cosmic string loops was proposed. This method produces identical results to standard methods, but has the advantage of requiring significantly less computation time. For this reason, we shall use this method when computing the SGWB spectrum in the small-loop regime.

In the small-loop regime, cosmic string loops live less than a Hubble time, $t_H=H^{-1}$. It is, therefore, reasonable to assume that their energy is radiated in GWs immediately after formation. This energy, however, is not radiated in a single frequency: as the loop size decreases, the GW frequency must increase. In this case, though, this occurs effectively immediately in the cosmological timescale. Thus, if the size of the loop at the moment of creation is $l(t)$, it radiates GWs with frequencies

\be
f>f_{\rm min}=\frac{2j}{l(t)}\frac{a(t)}{a_0}\,,
\ee
at the present time.

The distribution of the power radiated by small loops over the different frequencies is described by the following probability distribution function \cite{Sousa:2014gka}

\be
p(f)=p(l)\left|\frac{dl}{df}\right| \Theta(f-f_{\rm min})=\frac{f_{\rm min}}{f^2}\Theta (f-f_{\rm min})\,,
\label{pdf}
\ee
where $\Theta(f-f_{\rm min})=1$, for $f\ge f_{\rm min}$, and vanishes for all other $f$. Hence

\be
\left. \frac{d\rho_{\rm GW}}{dtdf}\right|_{\rm loops}=\left.\frac{d\rho}{dt}\right|_{\rm loops}\left(\frac{a(t)}{a_0}\right)^{4}\frac{f_{\rm min}}{f^2}\Theta(f-f_{\min})\,,
\ee
where $d\rho/dt |_{\rm loops}$ is given by Eq. (\ref{loss}). The spectral density of gravitational waves in the $j$-th mode of emission may, then, be computed as follows \cite{Sousa:2014gka}

\be
\Omega_{\rm gw}^j(f)=\frac{16\pi G}{3H_0^2 a_0^5}\int_{t_ {\rm min}}^{t_0}\left.\frac{d\rho}{dt}\right|_{\rm loops}\frac{2ja^5(t)}{\alpha f L(t)}dt \,,
\label{omegasmall}
\ee
where $t_{\rm min}$ is the time of creation of the loops that have $f_{\rm min}=f$.

\section{Cosmic Superstrings \label{superstrings}}

Cosmic superstrings and ordinary cosmic strings have different phenomenologies. First of all, there are two types of cosmic superstrings --- Fundamental strings, or F-strings, and $1$-dimensional Dirichlet branes, known as D-strings --- that have different tensions $\mu_F=g_s\mu_D$ (where $\mu_F$ and $\mu_D$ are, respectively, the tension of F- and D-strings and $g_s$ is the dimensionless string coupling). Moreover, cosmic superstrings may have a reconnection probability $\mathcal{P}$ that is significantly smaller than unity: it has been shown that $0.1\lesssim \mathcal{P} \lesssim 1$ for D-string interactions, while for F-string crossings $10^{-3}\lesssim \mathcal{P} \lesssim 1$ \cite{Jackson:2004zg}. Then, when two superstrings of the same type collide, they may either pass through each other without interacting or intercommute. As a result, these networks lose energy less efficiently and may, thus, be expected to have an energy density that is larger than that of ordinary cosmic string networks.

The most significant difference between super- and ordinary cosmic string networks results, however, from interactions between strings of different kinds. F- and D-strings do not intercommute: when an intersection occurs, the strings coalesce along their length and bind, forming a new type of string. As a matter of fact, this process occurring recursively may lead to the formation of bound states of $p$ F-strings and $q$ D-strings, known as $(p,q)$-strings, with a tension of

\be
\mu(p,q)=\mu_F \sqrt{p^2+q^2/g_s^2}\,,
\ee
where $p$ and $q$ are coprime. One may then expect the natural evolution of cosmic superstring networks to lead to an infinite hierarchy of cosmic superstrings with increasing tension, forming highly entangled networks with Y-type junctions in which three types of strings meet.

In Refs. \cite{Avgoustidis:2007aa,Avgoustidis:2009ke}, the VOS model for ordinary cosmic strings was extended to allow for the description of multi-tension networks with junctions. This was done by assuming that networks of cosmic strings of different types have different energy densities, $\rho_i$, and that they should be, as a consequence, characterized by different characteristic lengths $L_i$:

\be
\rho_i=\frac{\mu_i}{L_i^2}\,.
\ee
Here, the subscript $i$ is used to refer to the $i$-th type of string. Whenever two string species $i$ and $j$ interact, a portion of their length is used to create a new segment of type $k$. This results in a transference of energy from the $i$ and $j$ networks into the network of $k$ strings. This energy transfer may be described using the function

\be
\mu_k D_{ij}^k=\mu_k {\tilde d}_{ij}^{k} \frac{\vv_{ij}\ell_{ij}}{L_i^2L_j^2}\,,
\ee
where ${\tilde d}_{ij}^k={\tilde d}_{ji}^k$ is a phenomenological parameter that describes the efficiency of the junction formation mechanism in collisions of $i$ and $j$ strings, $\vv_{ij}=\sqrt{\vv_i^2+\vv_j^2}$, $\vv_i$ is the RMS velocity of the string of type $i$, and $\ell_{ij}$ is the average length of the new segment of string ($\ell_{ij}$ cannot be larger than the smallest of the characteristic lengths of the interacting networks). Here, we shall make the choice

\be
\ell_{ij}^{-1}=L_i^{-1}+L_j^{-1}\,,
\ee
introduced in \cite{Avgoustidis:2007aa}.

Note that, since in general $\mu_k \neq \mu_i+\mu_j$, there is some excess energy left behind in the junction formation process. One of the open questions about cosmic superstring dynamics is what happens to this excess energy: it may either be acquired by the new segment as kinetic energy, or radiated away (e. g. due to the production of microscopic F-strings \cite{Avgoustidis:2009ke}). One may take these possibilities into account by including in the velocity evolution equation, for each possible interaction, an acceleration term of the form

\be
\left.\frac{d\vv_k}{dt}\right|_{juncs}=\left(1-\vv_k^2\right)BD_{ij}^k\frac{\mu_i+\mu_j-\mu_k}{\mu_k}\frac{L_k^2}{\vv_k}\,,
\ee
where $0\le B\le1$ is a parameter that sets the portion of energy that is radiated away in the junction formation process: for $B=0$, all the excess energy is radiated away, while, for $B=1$, it is absorbed as kinetic energy.

The evolution equations for the characteristic length and RMS velocity of a network of superstrings of type $i$  are, then, of the form \cite{Avgoustidis:2007aa,Avgoustidis:2009ke}

\bq
\dtot{\vv_i}{t} & = & \left(1-\vv_i^2\right)\left[\frac{k_i(\vv_i)}{L_i} -\right.\nonumber\\
& - &  \left. 2H\vv_i+B\sum_{b,a\le b}D_{ab}^i\frac{\mu_a+\mu_b-\mu_i}{\mu_i}\frac{L_i^2}{\vv_i}\right]\,,\\
\dtot{L_i}{t} & = & HL_i\left(1+\vv_i^2\right)+\nonumber\\
 & + & \frac{1}{2}{\tilde c}_i \vv_i+\frac{1}{2}\left(\sum_{a,k}D_{ia}^k-\sum_{b,a\le b}D_{ab}^i\right)L_i^3\,,
\eq
where ${\tilde c}_i$ is the self-intersection (or loop-chopping) efficiency parameter for strings of type $i$, and we have assumed that $k_i(\vv_i)\equiv k(\vv)$ for all types of strings.

Note that when a $(p,q)$-string meets a $(p',q')$-string, there are two possible outcomes: either a $(p+p',q+q')$-string or a $(p-p',q-q')$-string is formed, depending on the relative velocity of the strings and on the angle of incidence. However, as $p$ and $q$ (and/or $p'$ and $q'$) increase, the probability of an additive process occurring decreases. The creation of heavy string types is thus not favored and one should expect the energy density of strings to decrease steeply as their tension increases. As matter of fact, it has been shown in \cite{Avgoustidis:2009ke,Pourtsidou:2010gu} that, in general, it suffices to consider the three lightest string types: $(1,0)$, $(0,1)$ and $(1,1)$. For the remainder of this paper, we shall consider only these three types of strings, which will be labeled $1$, $2$, and $3$, respectively.

To fully determine the cosmological evolution of cosmic superstrings, one thus needs six parameters: three self-interaction coefficients --- $\cc_1$, $\cc_2$, and $\cc_3$ --- and three cross-interaction coefficients --- $\dd_{12}^3$, $\dd_{13}^2$, and $\dd_{23}^1$. These parameters are determined by the microphysical intercommuting probabilities $\mathcal{P}_{ij}$ (that describe the interactions of strings of types $i$ and $j$ at a quantum level). The value of these parameters was computed in Ref. \cite{Pourtsidou:2010gu}. In general, one may expect that

\be
\cc_i=\cc\,\mathcal{P}_{i}^{\gamma}\,,
\label{cprob}
\ee
where $\cc$ is the loop-chopping efficiency of ordinary cosmic strings and $\pr_i\equiv \pr_{ii}$ is the intercommuting probability of cosmic strings of type $i$. In the context of a one-scale model, one would expect $\gamma=1$. Note however that Nambu-Goto simulations  of cosmic string network evolution with reduced intercommuting probability \cite{Avgoustidis:2005nv} indicate that the amount of small-scale structure on strings increases significantly as $\pr_i$ is reduced and, as consequence, a smaller value of $\gamma=1/3$ is observed. On the other hand, flat spacetime simulations \cite{Sakellariadou:2004wq} indicate a different exponent: $\gamma=1/2$. In Ref. \cite{Pourtsidou:2010gu}, when computing the self- and cross-interaction parameters, the authors assumed that $\gamma=1/3$ (and, thus, by using the values obtained therein, so do we). There, it is also assumed that the $\dd_{ij}$ parameters have the following dependence on the microphysical probabilities

\be
\dd_{ij}^k=d_{ij}^k\, S_{ij}^k\,,\quad\mbox{with}\quad d_{ij}^k=\kappa\,\mathcal{P}_{ij}^{1/3}\,,
\ee
where $\kappa\sim 1$ and $S_{ij}^k$ describes the conditional probability of a segment of type $k$ being produced in a collision of strings of types $i$ and $j$, given that an interaction has occurred. The formation of a zipper of type $k$ in collisions of  $i$ and $j$ strings is subject to kinematic constraints \cite{Copeland:2006eh,Copeland:2007nv,Salmi:2007ah,Avgoustidis:2009ke} (in particular, due to the existence of an additive and subtractive channel). The factors $S_{ij}^k$ account for the reduction in the average values of $\dd_{ij}^k$ coefficients that results from ``removing" all kinematically forbidden interactions (that realistically do not occur). The resulting values of the coefficients --- recorded in Table \ref{coeff} --- are also dependent on the string theory model that originates the superstrings: they depend not only on the string coupling $g_s$, but also on the volume of the compact extra dimensions that strings may explore. Here, as in \cite{Pourtsidou:2010gu}, this is parameterized by a volume factor $w\equiv V_{\rm min}/V_{FF}$, where $V_{\rm min}$ is the string scale volume and $V_{FF}$ is the effective volume the strings can explore.

Once the kinematic constraints on the junction formation process are taken into account (using the values in Table \ref{coeff} for the self- and cross-interaction parameters), linear scaling regimes seems to be a generic attractor solution of the generalized VOS equations for cosmic superstrings, both in radiation and matter dominated universes. Although, it is currently not clear what happens to the excess energy liberated in the binding process, these linear scaling regimes seem to occur, in general, for all $0\le B\le 1$. As pointed out in \cite{Avgoustidis:2009ke}, changing the value of $B$ does not affect the overall dynamics of the network drastically: it only seems to affect the heavier modes.

\begin{table*}[]
\centering
\begin{tabular}{|c|c|c|c|c|c|c|c|l|c|c|c|c|c|c|c|c|}
\cline{1-8} \cline{10-17}
$w$                & $g_s$ & $\cc_1$ & $\cc_2$ & $\cc_3$ & $\dd_{12}^3$ & $\dd_{13}^2$ & $\dd_{23}^1$ & \multirow{8}{*}{} & $w$                  & $g_s$ & $\cc_1$ & $\cc_2$ & $\cc_3$ & $\dd_{12}^3$ & $\dd_{13}^2$ & $\dd_{23}^1$   \\ \cline{1-8} \cline{10-17}
\multirow{7}{*}{1} & 0.04  & 0.02    & 0.13    & 0.13    & 0.05       & 0.08       & 0.55       &                   & \multirow{7}{*}{0.1} & 0.04  & 0.01    & 0.13    & 0.13    & 0.05       & 0.07       & 0.55       \\ \cline{2-8} \cline{11-17} 
                   & 0.1   & 0.03    & 0.16    & 0.16    & 0.04       & 0.11       & 0.62       &                   &                      & 0.1   & 0.02    & 0.16    & 0.16    & 0.04       & 0.1        & 0.62       \\ \cline{2-8} \cline{11-17} 
                   & 0.2   & 0.05    & 0.19    & 0.19    & 0.03       & 0.14       & 0.63       &                   &                      & 0.2   & 0.02    & 0.19    & 0.19    & 0.03       & 0.13       & 0.63       \\ \cline{2-8} \cline{11-17} 
                   & 0.3   & 0.07    & 0.20    & 0.20    & 0.03       & 0.16       & 0.61       &                   &                      & 0.3   & 0.03    & 0.20    & 0.20    & 0.02       & 0.14       & 0.61       \\ \cline{2-8} \cline{11-17} 
                   & 0.5   & 0.10    & 0.21    & 0.21    & 0.02       & 0.21       & 0.54       &                   &                      & 0.5   & 0.05    & 0.20    & 0.21    & 0.01       & 0.15       & 0.54       \\ \cline{2-8} \cline{11-17} 
                   & 0.7   & 0.12    & 0.22    & 0.22    & 0.02       & 0.26       & 0.49       &                   &                      & 0.7   & 0.06    & 0.15    & 0.22    & 0.01       & 0.17       & 0.39       \\ \cline{2-8} \cline{11-17} 
                   & 0.9   & 0.15    & 0.22    & 0.22    & 0.02       & 0.31       & 0.45       &                   &                      & 0.9   & 0.07    & 0.12    & 0.21    & 0.01       & 0.20       & 0.31           \\ \cline{1-8} \cline{10-17} 
\end{tabular}
\caption{The auto- and cross-interaction parameters, $\cc_i$ and $\dd_{ij}^k$, for the three lightest string types, for different values of $g_s$ and $w$. These coefficients were computed in Ref. \cite{Pourtsidou:2010gu}}
\label{coeff}
\end{table*}

\section{SGWB generated by cosmic superstring \label{sgwbsuper}}

Cosmic superstrings, despite having a reduced intercommuting probability, are still expected to produce a copious amount of loops. The cosmological evolution of superstrings results, as we have seen, in the creation of a hierarchy of string networks with increasing tension. One thus expects the evolution of cosmic superstring networks to give rise to several populations of loops of strings of different types. These different loop populations are (mostly) independent, and, thus, their contributions to the SGWB may be calculated separately.

Usually, when forecasting the SGWB spectrum generated by cosmic superstrings, it is common to merely compute the spectrum generated by ordinary cosmic strings with reduced intercommuting probability (see e. g., \cite{Damour:2004kw,Abbott:2009rr,Sanidas:2012ee,Binetruy:2012ze,Kuroyanagi:2012wm,Aasi:2013vna,Henrot-Versille:2014jua}). These studies only take into account the contributions of the lightest string type to the SGWB, and do not take into account the effect of the energy loss caused by junction formation. At a first glance, considering only the lightest string species --- the F-strings --- seems to be a reasonable assumption: heavier strings have a larger characteristic length and, consequently, their rate of loop production, as eq. (\ref{loss}) indicates, is expected to be significantly smaller than of lighter stringer. This means that one should expect, in general, the amplitude of the SGWB produced by the heavier strings to be significantly smaller. Note however that, since the characteristic length of the different species is different (and, thus, they emit gravitational waves with different frequencies), there may be signatures of the heavier strings on the total SGWB spectrum. In this section, we investigate this possibility.

\subsection{Can the VOS model for ordinary strings be used to describe F-strings (at least to some extent)?\label{voscomp}}

The VOS model for ordinary cosmic strings may be straightforwardly re-parametrized to describe string networks with reduced intercommuting probability: one simply needs to adjust the energy loss parameter according to Eq. (\ref{cprob}). Including the effect of junctions, however, is not as easy: it would require a numerical fitting of the $\cc$ parameter. Note however that, when the energy lost by F-strings in the junction formation process is negligible --- or, equivalently, if D-strings are considerably heavier than F-strings ($g_s\ll 1$) --- this re-scaling of the $\cc$ parameter should suffice to obtain a correct description of the dynamics of F-strings using the ordinary VOS model.

In this case, one may easily estimate the effect that the reduction of $\mathcal{P}$ has on the radiation era amplitude of the SGWB. It has been demonstrated in \cite{Avelino:2012qy} that, during the radiation era, weakly interacting networks experience a linear scaling regime of the form:

\be
\xi=\sqrt{2}\cc\,,\quad\mbox{and}\quad\vv=\frac{1}{\sqrt{2}}-\frac{\pi}{12}\cc\,,
\ee
for $G\mu\ll\cc\ll 1$. It is, therefore, straightforward to show that, during the radiation era, the rate of loop production scales as

\be
\dtot{n_c}{t}\propto \cc^{-3} \propto \mathcal{P}^{-1}\,,
\ee
while the frequency of the emitted radiation suffers a shift of

\be
f \propto \cc^{-1} \propto \mathcal{P}^{-1/3}\,.
\label{freqshift}
\ee
The amplitude of the flat portion of the SGWB consequently scales as

\be
\Omega_{\rm GW}(f)\propto \cc^{-2} \propto \mathcal{P}^{-\frac{2}{3}}\,.
\label{omegashift}
\ee
Note however that the effect on the peak of the spectrum (generated in the matter era) is not trivial: since after the radiation-matter transition is triggered, the network can no longer be assumed to be in a linear scaling regime, finding the explicit dependence of $\sgwb(f)$ on $\pr$ is not possible. In this case, Eqs. (\ref{freqshift}) and (\ref{omegashift}) are approximately verified, however the increase in the amplitude of $\sgwb (f)$ may be smaller in the large-loop regime. So --- although the cosmological constraints on cosmic strings that result from probes that test the flat portion of the spectrum (for instance, ground-based interferometers) may be trivially extended for F-string --- the accurate computation of the constraints that result from the portion of the spectrum associated to the gravitational radiation produced during the matter era necessarily involves a full recomputation of the spectrum (even when considering the lightest strings only).

Notice also that the shift in frequency of the emitted gravitational radiation in Eq. (\ref{freqshift}) is often ignored in the literature, with authors considering simply an enhancement of the amplitude of $\sgwb(f)$ when $\pr$ is reduced. This shift necessarily occurs and it has a simple physical interpretation. The effect of reducing the intercommuting probability is to also reduce the amount of energy that is lost into loops. This leads to an increase in the cosmic string energy density and to the corresponding reduction of the characteristic lengthscale. One therefore should have production of loops with a smaller physical size --- even assuming the same $\alpha$ --- that emit gravitational waves with larger frequency. The position of the peak of the spectrum is altered by this increase in the frequency of gravitational waves and, thus, ignoring this effect necessarily introduces errors in the computation of observational constraints for cosmic string with reduced $\pr$. By using the VOS model, parameterized by $\cc$, to describe the cosmological evolution of the cosmic string networks, we shall take this effect into account throughout this paper.

\begin{figure*}
\centering
\includegraphics[width=5in]{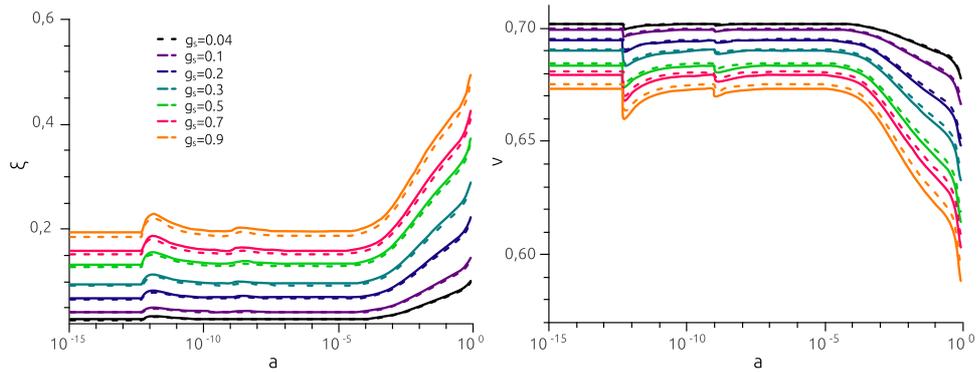}
\caption{The cosmological evolution of the RMS velocity and characteristic length of F-strings described by the VOS model for superstrings (solid lines), alongside that of cosmic strings described by the simplified model (dashed lines), for different values of $g_s$. Here, we have assumed that $w=1$ and $B=0$.}
\label{VOS}
\end{figure*}

Fig. \ref{VOS} shows the evolution of the RMS velocity and characteristic lengthscale of F-strings for different values of the string coupling $g_s$. There, we also plot the evolution of $\vv$ and $L$ for ordinary cosmic strings with the same (reduced) intercommuting probability. This is straightforwardly done by re-scaling the value of the loop-chopping parameter to that of the F-strings: $\cc=\cc_1$. Such procedure corresponds to what is usually done in the literature when dealing with the SGWB generated by cosmic superstrings and we shall refer to this model as the simplified model. In this computation and throughout the remainder of this paper, we take $H_0=100h\,\,{\rm km s}^{-1}\,{\rm Mpc^{-1}}$, with $h=0.673$, for the current value of the Hubble parameter, a fractional cosmological constant density of $\Omega_{\Lambda}^0=0.685$, and a radiation-matter equality redshift of $z_{\rm eq}=3393$, as indicated by the Planck 2015 data \cite{Ade:2015xua}. Moreover, we shall also include the dynamical effects associated with the alteration of the number of degrees of freedom caused by the annihilation of massive particles during the radiation era. This causes a sudden change on $H$ that affects temporarily cosmic string dynamics. Here, this effect on $H$ is modeled as in \cite{kolb}. As this figure clearly illustrates, the fit of the simplified model becomes increasingly poor as $g_s$ increases, despite providing a pretty good description of F-strings for $g_s \lesssim 0.1$. This difference is necessarily amplified in the SGWB spectrum, since the amplitude of $\sgwb (f)$ has a strong dependence on $L$ and $\vv$. One may therefore conclude that, in most instances, ignoring the heavier strings may introduce error in the computation of the SGWB generated by cosmic superstring networks. This figure also illustrates the effects of particle annihilation on the evolution of cosmic strings: when these events occur, the network is temporarily `knocked out' of the scale-invariant regime. After each event, the network then slowly evolves towards this regime again (which is, as we have seen, an attractor).

\subsection{Can the GW emitted by heavier strings really be neglected?\label{heavier}}

Heavier string species may affect the shape of the SGWB in two ways: they may not only affect the dynamics of F-strings --- which are, in general, the dominant contributors to the spectrum --- but also generate GWs themselves that contribute to this background. To investigate the relevance of this contribution, we compute the SGWB spectrum generated by the three lightest cosmic string types separately and add their contributions to each particular frequency. In these calculations, we shall assume that the three loop species are characterized by the same loop size parameter $\alpha_1=\alpha_2=\alpha_3=\alpha$, and that the gravitational emission efficiency parameter, $\Gamma$, is equal for all of them. Moreover, we shall start by studying loops created at two typical sizes (motivated by simulation results): large loops, which we shall assume to be characterized by $\alpha=1/20$, and small loops, with $\alpha=G\mu_1$. We will discuss the intermediate scales when appropriate. We shall also somewhat relax the assumption that all loops have the same $\alpha$ later in this section. Finally, in this subsection, we choose the fiducial set of parameters $G\mu_F=10^{-9}$, $B=0$, and $w=1$. The effect of the variation of the cosmic string tension on the shape and amplitude of the spectrum was discussed elsewhere (see e.g. \cite{Sanidas:2012ee,Sousa:2013aaa}). We shall discuss the impact of $B$ and $w$ on the shape of the spectrum in the next subsection.

\begin{figure*}
\centering
\includegraphics[width=7in]{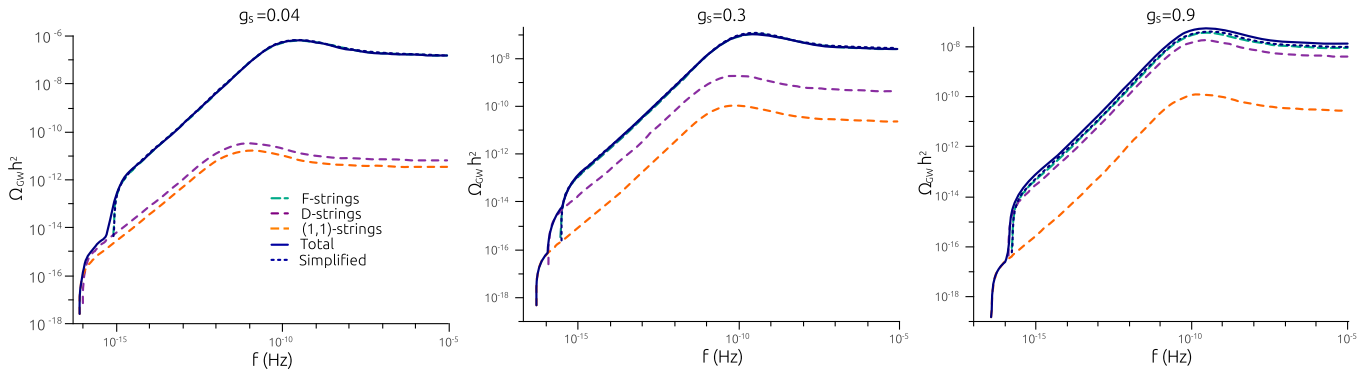}
\caption{The Stochastic Gravitational Wave Background generated by cosmic superstrings (solid lines) characterized by $G\mu_F=10^{-9}$, $B=0$ and $w=1$, for different values of the fundamental string coupling $g_s$, in the large loop regime ($\alpha=1/20$). Here, we also plot the SGWB generated by each of the string species (dashed lines) and that of cosmic superstrings described by the simplified model (dashed line).}
\label{large}
\end{figure*}

Fig. \ref{large} shows three examples of the SGWB generated by cosmic superstring networks in the large loop regime (solid lines), for different values of $g_s$ and $w=1$. For each of these, we also plot the SGWB generated by cosmic superstrings described by the simplified model (dotted line). One may notice that, for large loops, the simplified model provides a fairly good description of most of the SGWB spectrum generated by cosmic superstrings for sufficiently low $g_s$ (if one takes into account, as we do here, the frequency increase that accompanies the reduction of $\pr$). Note however that these spectra are not identical: there is a small difference in the amplitude throughout (that is not apparent in the figures because the $\sgwb (f) h^2$ axis spans 12 orders of magnitude). This similarity in the spectrum necessarily means that one may not be able to distinguish between cosmic superstrings and ordinary strings that have a reduced intercommuting probability due to non-standard interaction mechanisms (e.g. friction caused by interaction with other cosmological components, the existence of conserved currents, etc.) if a signal is detected. This difficulty may even arise if $g_s$ is large: one may always fit the superstring spectrum using the simplified model by using different $G\mu$ and $\cc$ values. This degeneracy is only broken if one looks to small enough frequencies: in that range, there exist clear signatures of heavier modes (that emit at lower frequencies because their characteristic length is larger). Detecting them would allow us, at least in principle, to obtain more information not only about the string coupling but also about the underlying string theory. Note, however, that the expected frequency range and the magnitude of the predicted signatures may be beyond the reach of current gravitational wave experiments. This discussion applies for all values of $\alpha$ in the large loop regime (with $\alpha\gg G\mu$) --- although, as $\alpha$ increases, the peaks generated by the different string types get closer in amplitude

\begin{figure*}
\centering
\includegraphics[width=7in]{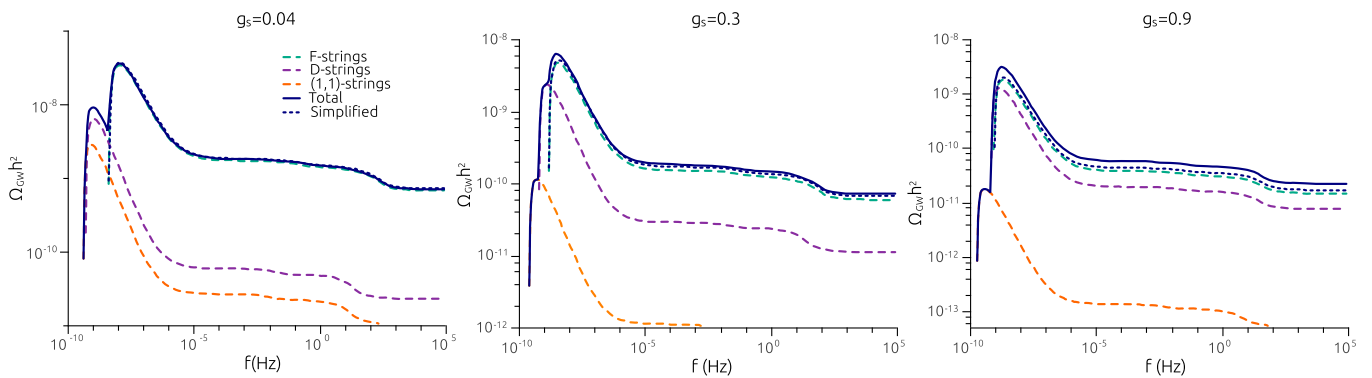}
\caption{The Stochastic Gravitational Wave Background generated by cosmic superstrings (solid lines) characterized by $G\mu_F=10^{-9}$, $B=0$ and $w=1$, for different values of the fundamental string coupling $g_s$, in the small-loop regime (with $\alpha=10^{-9}$). Here, we also plot the SGWB generated by each of the string species (dashed lines) and that of cosmic superstrings described by the simplified model (dashed line).}
\label{small}
\end{figure*}

The same is not true in the small-loop regime. In Fig. \ref{small}, we plot the total spectrum of gravitational waves generated by cosmic superstrings with different $g_s$ values (solid lines), alongside that of superstrings described by the simplified model, in the small-loop regime. Although, as previously discussed, the fit of the simplified model to the SGWB generated by F-strings may be considered fairly good in most instances, the signatures of heavier modes on the peak of the spectrum are, in this case, much more prominent. As a matter of fact, their detection is indeed conceivable in this instance. Furthermore, the type of signature left by the heavier modes is dependent on the relative tension of the different species (or, equivalently, the value of $g_s$). For $g_s\sim 1$, F- and D-strings have comparable tensions and, thus, they contribute mostly to the same frequency range and their peak has a similar location. The SGWB signal of F-strings is therefore enhanced by the contribution of D-strings --- which makes the simplified model a poorer fit in this case. Evidence of the existence of $(1,1)$-strings may be found in the lowest frequency range in the form of a small ``bump''. If, however, $g_s\ll 1$, type 2 and 3 strings have similar tensions and, thus, their contribution to the spectrum may be observed in the form of a larger ``bump'' --- almost a second peak --- in the low frequency range. Although, at a first glance, these two situations may look similar, in the case of $g_s\ll 1$, the signature of heavier modes is significantly more prominent and there is a tenuous signature of the type 3 strings (caused by the fact that the peak of their SGWB spectrum is not exactly coincident with that of type 2 strings). Finally, for intermediate values of $g_s$, we may have a situation in which signatures of all three types of strings are observable (since their peaks happen at different frequencies). In this case, therefore, there is a significant broadening of the peak of the observed spectrum. In either case, the signatures of the heavier modes seem to be within the sensitivity window of upcoming gravitational wave detectors and, therefore, a direct detection of these features is conceivable. Having said this, notice that we have only considered the three lightest types of strings. Heavier modes may also leave specific signatures, albeit at even lower frequencies and with significantly fainter amplitude.

\begin{figure*}
\centering
\includegraphics[width=7in]{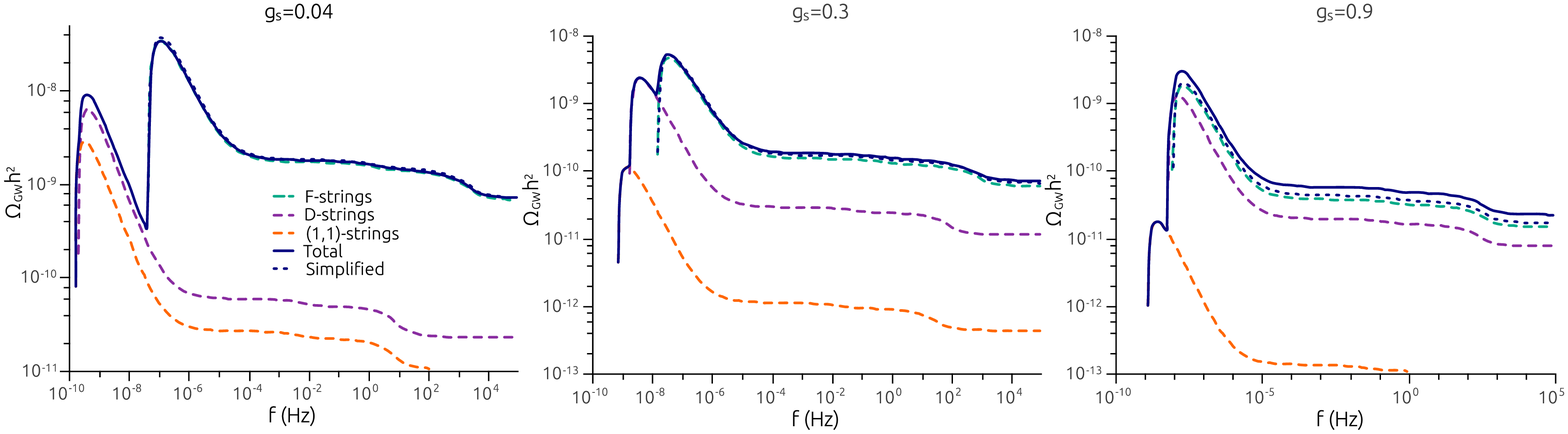}
\caption{The Stochastic Gravitational Wave Background generated by cosmic superstrings (solid lines) characterized by $G\mu_F=10^{-9}$, $B=0$ and $w=1$, for different values of the fundamental string coupling $g_s$, for which loop production happens close to the gravitational backreaction scale ($\alpha_i=G\mu_i$). Here, we also plot the SGWB generated by each of the string species (dashed lines) and that of cosmic superstrings described by the simplified mode (dashed line).}
\label{backreaction}
\end{figure*}

As to relaxing the assumption that all string species produce loops with the same characteristic loop size parameter, one may clearly construct spectra with completely different shapes and with different string types providing the dominant signatures. Note however that there is no physical evidence supporting loop production occurring at significantly different scales for different string types. The situation in which the loop size is close to (but slightly smaller than) the gravitational backreaction scale --- i.e., $\alpha_i\sim G\mu_i$ --- , however, is of physical interest: (standard) Nambu-Goto simulations indicate that $90\%$ of the energy lost into loop production is in the form of loops created at the gravitational backreaction scale. In Fig \ref{backreaction}, we plot the SGWB generated by cosmic superstring networks that have $\alpha_i=G\mu_i$ for three values of the fundamental string coupling ($g_s$). As one might expect, this situation is quite similar to that of the small-loop regime when all $\alpha$'s are equal discussed previously in this section. Note however that, in this situation --- wherein strings with larger tension also have a correspondingly larger $\alpha$ --- the signatures of heavier strings are more prominent and may be easier to detect.

\subsection{What can we realistically expect for cosmic superstrings?\label{realistic}}

The shape of the SGWB generated by cosmic superstrings is, as is also the case for ordinary cosmic strings, not only dependent on the macroscopic properties of the network ($G\mu_i$, $L_i$ and $v_i$) but also on the size and emission spectrum of cosmic string loops. There are several aspects of the evolution of cosmic superstring networks that are not entirely understood wet. However, there are several studies of cosmic superstring dynamics in the literature that may help us shorten the parameter space.

In \cite{Avgoustidis:2005nv}, it has been demonstrated, using Nambu-Goto simulations of cosmic string dynamics, that there is a build up of small-scale structure on cosmic strings as the reconnection probability is reduced. It was also suggested therein that, due to this build up of small-scale structure, the reduction of $\pr$ should not be expected to have a significant impact on the production of small loops.  The production of large loops, however, should be suppressed in cosmic string networks with reduced intercommuting probability \cite{Avgoustidis:2005nv}. As we have discussed, recent simulations of Nambu-Goto cosmic strings with $\pr=1$ indicate that only $10\%$ of the energy lost as a result of string intercommutation is in the form of large loops. For cosmic superstrings, which may have a significantly reduced intercommuting probability, one would then expect the fraction of energy lost into large loops to be significantly reduced, and, therefore, that their contribution to the SGWB would be subdominant (if not negligible). It is therefore reasonable to assume that all loops (from all different cosmic string species) are created at the gravitational backreaction scale. 

Moreover, the existence of extra dimensions that the string may explore decelerates the strings in the non-compact dimensions. As a consequence, the formation of cusps --- in which strings are locally ultra-relativistic --- is significantly suppressed \cite{O'Callaghan:2010sy,Avgoustidis:2012vc}. The authors found that the formation of `near cusp events' (wherein the velocity of the string is locally smaller) may still occur, but, in this instance, their gravitational wave emission is significantly weaker. Kinks, on the other hand, are produced as a result of intercommutation and, therefore, should also be a generic feature of cosmic superstrings \cite{O'Callaghan:2010hq}. As a matter of fact, it has been shown in \cite{Binetruy:2010bq,Binetruy:2010cc} that kinks proliferate on superstrings due to the presence of Y-type junctions: when a kink, as it is propagating on a string, encounters a junction, it is reflected and it gives rise to two daughter kinks in the other two connecting strings. This effect could be dramatic in superstring loops with junctions --- in that case, the number of kinks on the loop would grow exponentially; however, when dealing with small loops, it is highly unlikely that such loops would be chopped off of the network. Loops with junctions may also result from the collision of two loops of strings of different types. In this case, however, the newly formed string segment unzips, leading to the separation of the loops \cite{Firouzjahi:2009nt}.

It seems therefore that, in realistic scenarios, one may expect the natural evolution of cosmic superstring networks to give rise to a large population of small and kinky loops, created at the gravitational backreaction scale --- which happens to be the case in which the signatures of heavier strings are more prominent.

\begin{figure*}
\centering
\includegraphics[width=7in]{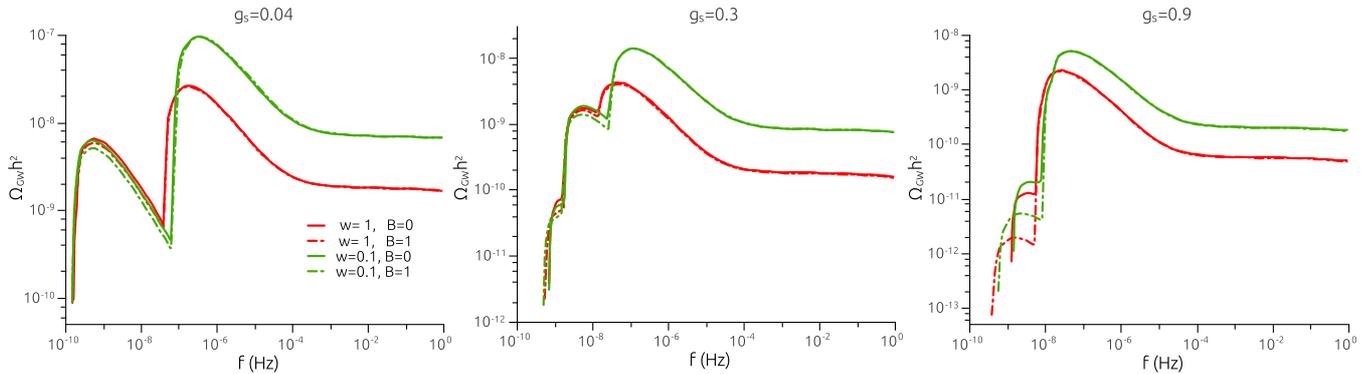}
\caption{The Stochastic Gravitational Wave Background generated by realistic cosmic superstrings networks --- with $G\mu_F=10^{-9}$, $\alpha_i=G\mu_i$, $q=2$, and $n_s=10^3$ --- for different values of the fundamental string coupling $g_s$. Here, we plot this spectra for two values of $w$ --- $w=1$ (red line) and $w=0.1$ (green line) --- and for $B=0$ (solid lines) and $B=1$ (dashed lines).}
\label{real}
\end{figure*}

In Fig. \ref{real}, we plot the SGWB generated by kinky cosmic string loops, whose size is set by the gravitational backreaction scale, for different values of $g_s$. We have assumed a spectral index $q=2$ and considered modes of emission up to $n_s=10^3$ (which is the saturation mode for loops with kinks \cite{Sanidas:2012ee,Sousa:2013aaa}). In this figure, we also relax the assumption that $w=1$ and $B=0$. The effect of including higher modes of emission in the shape of the peak of the SGWB spectrum is well known: it causes a slight decrease in its maximum amplitude that is accompanied by a broadening of the peak. Fig. \ref{real} shows that the inclusion of higher modes does not alter the general conclusions of the previous subsection: heavier strings do leave clear detectable signatures on the SGWB generated by cosmic superstrings. Thus, if cosmic superstrings are present, one should, in principle, be able to detect these signatures that distinguish them from ordinary cosmic strings. Furthermore, Fig. \ref{real} allows us to understand the effect of the existence of extra-dimensions on the shape of the spectrum: if the strings can move in more than $3+1$-dimensions (or if $w<1$), they are more likely to miss each other and avoid collision. This is the equivalent of having a reduction of the effective probability of reconnection and of having a less efficient energy loss mechanism. As a consequence, there is an enhancement of the string energy density and a resulting increase in the energy density of gravitational waves. This increase is particularly evident on type 1 strings, since heavier strings explore a smaller fraction of extra dimensions \cite{Jackson:2004zg,Pourtsidou:2010gu}.

Finally, as discussed in Sec \ref{superstrings}, changing the value of $B$ only seems to affect significantly the heavier cosmic strings. If one allows the energy left out in the junction formation process to be absorbed by the new segment, there is an increase in the RMS velocity of the corresponding string type and a consequent increase in its characteristic length. The effect of these changes is to lower the amplitude of the signatures generated by the heavier strings on the SGWB spectrum (see Eq. (\ref{nc})). This decrease is larger for increasing $g_s$, however, in either case, it is not sufficient to render these signatures unnoticeable.

\section{Observational Constraints on Cosmic Superstrings\label{obs}}

As previously discussed, the state of the art on cosmic superstrings seems to indicate that their cosmological evolution results in the copious production of loops with kinks whose size is set by the gravitational backreaction scale. In this section, we shall use the current observational constraints on the amplitude of the SGWB to derive constraints on the tension of fundamental strings, $G\mu_F$.

\subsection{CMB constraints\label{cmbsec}}

In the case of cosmic string networks that produce small loops --- which is the case when $\alpha\sim G\mu$ --- , indirect CMB constraints often provide the strongest limitis on the SGWB generated by cosmic strings. These indirect CMB constraints do not come from constraints on the B-mode polarization of CMB (which probe gravitational waves directly), but are inferred from the constraints on the number of relativistic degrees of freedom. The existence of a gravitational wave background in the early universe --- which behaves as a free-streaming gas of massless particles (as do neutrinos) --- necessarily affects the CMB and matter power spectra \cite{Smith:2006nka}. Any deviations of the observed effective number of relativistic degrees of freedom from the predicted value would then indicate the presence of additional relativistic radiation. Current Planck data, combined WMAP, SPT and ACT data, as well as Baryon acoustic oscillation and lensing data, allowed the authors of \cite{Henrot-Versille:2014jua} to derive, at a $95\%$ confidence level, the following upper limit on the total energy density of gravitational waves, in units of critical density, created until the time of decoupling, $t_d$, as would be observed today

\be
\left.\Omega(t_d)_{\rm GW}^{\rm Total}\right|_{t_0}=\int_{t_i}^{t_d}\sgwb(f)d(\ln{f})<3.8\times 10^{-6} h^{-2}\,,
\label{CMBlimit}
\ee
where $t_i$ is the time instant in which when the emission of gravitational waves is initiated. For cosmic strings, one should expect $t_i\sim t_f\sim t_{pl}(G\mu)^{-2}$ \cite{vilenkin2000cosmic}, where $t_f$ is the end of the friction-dominated era (during which the movement of strings is heavily damped) and $t_{pl}$ is the Planck time. Note that this constraint leaves out most of the gravitational waves created during the matter era, during which the peak of the spectra (and thus most of the signatures of heavier string types) are created. However, signatures of type 2 and 3 strings are still present in the low frequency range. Note also that big bang nucleosynthesis (BBN) allows us to put a constraint of similar magnitude on $\left.\Omega(t_{BBN})_{\rm GW}^{\rm Total}\right|_{t_0}h^2<8.1\times 10^{-6}$ \cite{Cyburt:2004yc} (where $t_{BBN}$ is the time in which big bang nucleosynthesis occurs). However since, in this case, only gravitational waves emitted until $t_{BBN}$ are constrained, this would result necessarily in weaker constraints. Having said this, we shall point out that when addressing the indirect CMB  or BBN constrains on SGWBs it is essential to include the effects of the annihilation of massive particles in the cosmological history. As we have shown in Sec. \ref{voscomp}, these cause a significant reduction of the amplitude of the gravitational wave power spectrum created early in the radiation era and, since this reduction significantly affects the total energy density in gravitational waves, not including their effect may lead to a significant overestimation of the constraints on $G\mu_F$.

\begin{figure}
\centering
\includegraphics[width=2.5in]{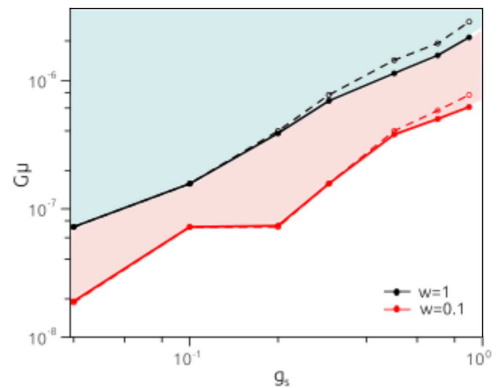}
\caption{Indirect CMB constraints on the tension of fundamental strings, $G\mu_F$, as a function of the string coupling $g_s$, for cosmic superstrings with $\alpha_i=G\mu_i$, $q=2$ and $n_s=10^3$. Black lines correspond to models for which $w=1$, while red lines represent the constraints for $w=0.1$. Solid lines represent the constraints derived using the velocity dependent model for cosmic superstrings and dashed lines correspond to those obtained using the simplified model. Shaded areas correspond to the excluded parameter regions. Here, we chose $B=0$.}
\label{CMB}
\end{figure}

In Fig. \ref{CMB}, we plot the observational constraints on the tension of fundamental strings, $G\mu_F$, as a function of the fundamental string coupling $g_s$, that result from the (indirect) CMB constraints on SGWBs (in Eq. (\ref{CMBlimit})) for kinky loops with $\alpha_i=G\mu_i$. Here, we have chosen $B=0$, since this choice leads to weaker constraints. Note that changing the value of $B$ has a negligible impact on the constraints on $G\mu_F$ for small $g_s$, and, as $g_s$ increases, the effect on the constraints also increases until it reaches about $5\%$ ($1\%$) for $g_s=0.9$ and $w=1$ ($w=0.1$) and $B=1$. In this figure, one may see that models with smaller $g_s$ are more tightly constrained by the CMB data than those with larger $g_s$. This happens because the total energy density in gravitational waves emitted by cosmic superstrings is larger for smaller intercommuting probabilities and, thus, for smaller $g_s$. For the same reason, the constraints on $G\mu_F$ for smaller $w$ are stronger than those with larger $w$: as Fig. \ref{CMB} shows, changing the value of $w$ from $1$ to $0.1$ leads to an increase of the area of parameter space that is excluded (represented in this picture by the pink shaded area). Note that cosmic superstrings may, in fact, have significant motion in the compact dimensions and they may be able to explore a larger fraction of their value than previously expected \cite{Avgoustidis:2012vc} --- instead of being localised in the internal dimensions, while fluctuating around a minimum of the potential well \cite{Jackson:2004zg}. In this case, one expects $w$ to assume values that are significantly smaller than unity, and the constraints on $G\mu_F$ to be stronger than the one presented here.

Indirect CMB constraints on SGWBs result in a conservative constraint on the tension of fundamental string of $G\mu_F<2.9\times 10^{-6}$. This constraint is conservative in the sense that it is the maximum tension allowed by CMB data independently of the value of $g_s$ and $w$. This constraint is approximately two orders of magnitude weaker than the constrains on cosmic superstrings that result from primary CMB anisotropies \cite{Charnock:2016nzm}: $G\mu_F<2.8\times 10^{-8}$. Note however that the constraints for $w=0.1$ and $g_s=0.04$ are slightly stronger: $G\mu_F<1.9\times 10^{-8}$. This clearly shows that indirect CMB constraints on the SGWB generated by cosmic superstrings have the potential of increasing the current exclusion region of parameter space: for instance, they yield stronger constraints for small enough $g_s$ and/or small $w$.

Finally, this figure also reveals the effect that using the simplified model to describe cosmic superstring dynamics has on the derived observational constraints. For small $g_s$, the simplified model provides a good estimate of the observational constraints on $G\mu_F$. However, for increasing $g_s$, the quality of this approximation becomes increasingly poor: the estimated constraints on $G\mu_F$ may be weaker by as much as $25\%$ for $g_s=0.9$. Still, the simplified model may be used, in this case, to provide robust conservative observational constraints. Note however that indirect CMB probes --- which only probe the total energy density in gravitational waves created until the time of decoupling --- are not very sensitive to the errors introduced by the simplification since the dominant contributors are type 1 strings. As we shall see in the next subsection, direct gravitational wave probes sensitive to the frequencies at which the peaks of the spectra are located can be more affected by these errors.

\subsection{Pulsar Timing array constraints\label{pulsarsec}}

Pulsar timing arrays probe the existence of SGWBs in the low frequency range ($f\sim 10^{-9}-10^{-8}$) of the gravitational wave spectrum and, for this reason, they have the strongest constraining power on the cosmic string tension for networks that produce large loops. Note however that, given the high sensitivity of these experiments in this region, one should expect them to also provide relevant constraints on the tension of cosmic superstrings for which loop production happens at the gravitational backreaction scale.

Pulsar timing constraints on the stochastic gravitational wave background are often expressed in terms of the strain of the spectrum, $\hgw$ ---
usually modeled as a power law of the form

\be
\hgw(f)=A_{\rm GW}(\nu)\left(\frac{f}{{\rm yr}^{-1}}\right)^{\nu}\,,
\label{strain}
\ee
where $A_{\rm GW}$ is the characteristic strain of the spectrum --- which is related to the spectral density in gravitational waves by

\be
\sgwb(f)=\frac{2\pi^2}{3H_0^2}f^2\hgw^2\,.
\ee

Observational constraints on the strain amplitude are dependent on the spectral index of the SGWB, $\nu$. NANOGRAV 9-year data \cite{Arzoumanian:2015liz} places an upper limit of $A_{\rm GW}<1.5\times 10^{-15}$ (at a $95\%$ confidence level), at a frequency $f=1\,\,{\rm yr}^{-1}$, for supermassive black hole binaries with $\nu=-2/3$ (which are expected to be a significant source in this frequency range). However, they also provide an analytical fit for the upper limit of the strain amplitude for different values of the power spectral index: $A_{\rm GW}\propto 10^{-0.4(3-2\nu)}$.

For cosmic strings, it is often assumed that $\nu=-7/6$, following estimations of the strain of the SGWB generated by cuspy loops \cite{Damour:2004kw}. However, as pointed out in \cite{Sanidas:2012ee,Sanidas:2012tf} (and as one may realize by analyzing the shape of the typical cosmic string spectrum), the slope of the SGWB cannot be assumed to be constant. One thus expects the upper limits on the spectral density that result from NANOGRAV (and other pulsar timing arrays) to depend on the spectral index of the spectrum $d$ --- defined by assuming that locally $\sgwb\propto f^d$ --- at the reference frequency of $f=1\,\,{\rm yr}^{-1}\simeq32 nHz$:

\be
\sgwb(f=1\,\,{\rm yr}^{-1})h^2 < 4.15\times 10^{-10+0.8d}\,,
\ee
where $d=2(\nu+1)$. For $\nu=-2/3$ (or equivalently $d=-1/3$), this yields an upper limit on the SGWB of cosmic strings of $2.4\times 10^{-10}$ (which is the constraint quoted in \cite{Arzoumanian:2015liz}). However, since $d >-1/3$ in large extents of the spectrum, using simply this upper limit may lead to overestimation of the constraints on $G\mu_F$.

\begin{figure*}
\centering
\includegraphics[width=5in]{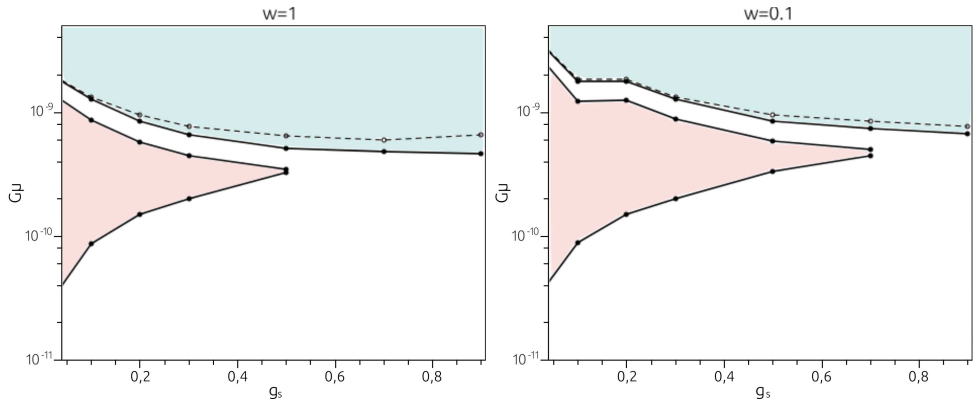}
\caption{NANOGRAV constraints on the tension of fundamental strings, $G\mu_F$, as a function of $g_s$, for cosmic superstrings with $\alpha_i=G\mu_i$, $q=2$, and $n_s=10^3$. The left panel corresponds to models with $w=1$ and the right panel to $w=0.1$. Solid lines represent the constraints derived using the velocity dependent model for cosmic superstrings and dashed lines to those obtained using the simplified model. The blue shaded area represents the exclusion region that results from F-strings, while the pink shaded area represents that resulting from heavier string types. Here, we chose B=0.}
\label{pulsar}
\end{figure*}

In Fig. \ref{pulsar}, we plot the constraints on the tension of fundamental strings, $G\mu_F$, as a function of the string coupling $g_s$, for $w=1$ and $w=0.1$. As this figure illustrates, for the case of cosmic superstring networks with $\alpha_i=G\mu_i$, the constraints that result from NANOGRAV at $f=1\,\,{\rm yr}^{-1}$ are significantly stronger than those resulting indirectly from CMB data: they allow us to establish a conservative limit on the tension of fundamental strings of

\be
G\mu_F<3.2\times 10^{-9}\,.
\label{conslimit}
\ee
This is about an order of magnitude stronger than the constraints that result from primary CMB anisotropies (in Ref. \cite{Charnock:2016nzm}) and it is merely the constraint that results from F-strings only.

Note however that these constraints go significantly beyond the conservative limit in Eq. (\ref{conslimit}). Pulsar timing arrays may be used to detect not only the SGWB generated by F-strings but also that of D-strings directly. The secondary exclusion region --- represented by the pink shaded area --- corresponds to the case in which the amplitude of the spectrum generated by D-strings --- sometimes in combination with that of $(1,1)$-strings when their frequency ranges coincide (for small $g_s$) --- exceeds the observational limits set by NANOGRAV. This region is only present for small enough $g_s$, when the spectra of type 1 and type 2 strings are well separated. As a matter of fact, a larger range of the tension of fundamental strings is excluded as $g_s$ decreases (for $g_s=0.09$, it spans about two orders of magnitude). When using the simplified model to describe the SGWB generated by superstrings (and neglecting the signatures of heavier modes), one looses information about this exclusion region. Moreover, as was the case for indirect CMB constraints, this figure also shows that using  simplified model is fairly successful in predicting the constraints on $G\mu_F$ that result from F-strings for small $g_s$ --- despite resulting in an underestimation by around $30\%$ of the constraint for large $g_s$. This model, however, cannot predict the secondary exclusion region and, thus, does not allow for a full exploration of the parameter space. These limitations of the simplified model apply not only to observational constraints resulting from pulsar timing arrays, but also to those that result from any direct probes of $\sgwb(f)$. Finally, note that although heavier string composites --- $(1,1)$-strings and beyond --- seem to be outside of the reach of current pulsar timing arrays, their detection with future experiments with larger sensitivity --- such as IPTA, SKA or eLISA --- is indeed conceivable.

\section{Discussion and Conclusions\label{conc}}

In this paper, we have studied in detail the SGWB generated by the three lightest types of cosmic superstrings. We have demonstrated that the inclusion of heavier string types is essential to make accurate predictions of the shape of the SGWB, and that type $2$ and type $3$ strings leave distinct signatures on the total spectrum on the low frequency range. The commonly used simplified model --- which approximates the spectrum of cosmic superstrings by that of ordinary strings with reduced intercommuting probability --- is, to some extent, successful in describing the spectrum generated by F-strings (particularly so if $g_s \ll 1$). However, by not including heavier strings (and their dynamical effects on type 1 strings), it introduces innacuracies in the computation of observational constraints on the tension of fundamental strings. These innacuracies are significant for direct gravitational wave detectors that probe the frequency range in which the peaks of heavier strings are located (such as pulsar timing arrays or space-borne observatories), since, as we have demonstrated, the simplified model does not predict the secondary exclusion region that results from heavier strings. However, since the constraints derived using the simplified model are, in general, weaker than those obtained using the VOS model for superstrings, they may be considered `safe' conservative bounds.

We have focused our study on the motivated scenario in which loops are kinky and their size is determined by the gravitational backreaction scale. This scenario is motivated theoretically and by numerical simulations of ordinary cosmic strings. However, since the properties of cosmic superstrings are significantly different from those of ordinary strings, merely extrapolating the results on loop production of ordinary cosmic strings may be overreaching. To accurately characterize loop production in cosmic superstring network evolution, one will need detailed numerical studies of cosmic superstring dynamics. Until those are performed, one should probably be conservative in the choice of range for the parameter $\alpha$ when deriving observational constraints on the fundamental string tension or predicting the parameter space available for future missions.

There are other aspects of loop production in cosmic superstrings that are not completely understood. For instance, the velocity-dependent model for cosmic superstrings (as well as that for ordinary strings) describes only the large scale dynamics of the networks, and does not provide a detailed model of the small-scale structure along the strings. Small-scale structure is expected to play a key role in loop formation --- particularly so in the case of small loops, whose size is expected to be dependent on the typical size of this structure. Understanding how small-scale structure evolves is essential to determine the length with which loops are created. Note however that studies performed using more complex models to describe cosmic string dynamics \cite{Quashnock:1990qy,Austin:1993rg} indicate that the characteristic lengthscale of small-scale structure scales with the characteristic length of the network. This justifies the choice of using $l=\alpha L$ even in the small-loop regime or when the size of the loop is determined by gravitational backreaction.

Another relevant question for the case of superstrings is whether the characteristic lengths of the different types of strings are adequate measures of their physical length. This is, in general, the case for ordinary cosmic strings (except in ultra-relativistic regimes that are only attained in contracting universes), however it may not be the case when zipper-type junctions are present. In general, one may expect the characteristic lengths of F- and D-strings to be an accurate measure of their physical length (since they form networks that may exist independently). However, the same may not be true for compound strings. The characteristic length --- which is a measure of the energy density of the network --- may assume a large value if the strings have a large physical length or if there exists a small number of segments with small physical length. Field theory simulations of cosmic superstrings seem to favor this latter scenario \cite{Urrestilla:2007yw,Lizarraga:2016hpd} (note however that these do not include the case in which $g_s\ll 1$). The physical length is the relevant length determining determining the size of loops. So, by using $L$, one may be overestimating the size of loops created by type 3 strings. Note, however, that this may be easily corrected by rescaling the value of $\alpha$ to a smaller value for compound strings. In the small-loop regime, this merely causes the SGWB generated by type 3 strings to shift towards higher frequencies and, thus, has no effect on the constraints resulting from the CMB. As to the constraints that result from the pulsar timing array data, this shift may put the spectrum of type 3 strings out of the probe's range. However, since their contribution to the spectrum is subdominant, the impact on the shape of the secondary exclusion region is expected to be limited too.

\acknowledgments

L.S. is supported by Funda{\c c}\~ao para a Ci\^encia e a Tecnologia (FCT) and and POPH/FSE through the grant SFRH/BPD/76324/2011. P.P.A. is supported by FCT through the Investigador FCT contract reference IF/00863/2012 and POPH/FSE (EC) by FEDER funding through the program Programa Operacional de Factores de Competitividade - COMPETE.  Funding of this work was also provided by the FCT grant UID/FIS/04434/2013.

\bibliography{SGWBsuper}

\end{document}